\documentclass[
aps,%
12pt,%
final,%
notitlepage,%
oneside,%
onecolumn,%
nobibnotes,%
nofootinbib,%
superscriptaddress,%
noshowpacs,%
centertags]%
{revtex4}
\usepackage{epsfig}
\usepackage{graphics}

\begin{document}

\title{ Direct $J/\psi$ and $\psi'$ hadroproduction via
fragmentation in collinear parton model and $k_T$-factorization
approach}

%
%
\author{\firstname{V.A.} \surname{Saleev}}
\email{saleev@ssu.samara.ru} \affiliation{Samara State University,
Samara, Russia and Samara Municipal Nayanova University, Samara,
Russia}
\author{\firstname{D.V.} \surname{Vasin}}
\email{vasin@ssu.samara.ru}
 \affiliation{Samara State University,
Samara, Russia}

%
%



\begin{abstract}
The $p_T$-spectra for direct $J/\psi$ and $\psi'$ in
hadroproduction at Tevatron energy have been calculated within the
framework of the NRQCD formalism and the fragmentation model in
the collinear parton model as well as in the $k_T$-factorization
approach. We have described the CDF data and obtained a good
agreement between the predictions of the parton model and the
$k_T$-factorization approach. We performed the calculations using
the relevant leading order in $\alpha_s$ hard amplitudes and
taking equal values of the long-distance matrix elements for the
both models.
\end{abstract}

\maketitle

\section{Introduction} During the years after the measuring of
charmonium productiom cross sections and polarization effects for
$J/\psi$ and $\psi'$ mesons at the Tevatron Collider \cite{1} the
phenomenology of a quarkonium production has a phase of intensive
developments. Nowadays, it is understand that the heavy quarkonium
production is a very complicated physical process needs many new
theoretical ideas. Starting from the Color Singlet Model \cite{2}
it was developed, so-called, the nonrelativistic QCD  (NRQCD)
formalism \cite{3} to describe the nonperturbative transition of
$Q\bar Q$-pair into a final heavy quarkonium.

 The perturbative
fragmentation functions for partons which split into heavy
quarkonia have been obtained \cite{4} within the framework of the
NRQCD formalism. It is supposed that the fragmentation model is
more adequate for the description of a quarkonium production at
the large quarkonium transverse momentum $p_T\gg M_{Q\bar Q}$ than
the fusion model.

In the charmonium production at the energy range of the Tevatron
Collider  we deal with the gluon distribution function from a
proton which is taken at a very small $x$ but the relevant
virtuality $\mu^2\sim M_{Q\bar Q}^2+p_T^2$ is large. In the region
under consideration the collinear parton  model can be generalized
within the framework of the $k_T-$factorization approach
\cite{5a,5b,5c,5d}. This fact leads to some interesting effects in
the quarkonium production at the high energies, which were
discussed ten years ago in Ref. \cite{6} and recently in Refs.
\cite{7,7a,7b,7c}.

In this paper we calculate the $p_T$-spectra of the unpolarized
direct $J/\psi$ and $\psi'$ mesons produced via the fragmentation
mechanism. In case of  direct $J/\psi$ and $\psi'$ it is supposed
that the production via the gluon fragmentation into the
color-octet $Q\bar Q$ [$^3S_1, \underline{8}$] state is a dominant
contribution \cite{8,9}. We compare the predictions which are
obtained as in the collinear parton model as in the
$k_T$-factorization approach. In both cases we performed
calculations with the hard amplitudes in the leading order of the
QCD running constant $\alpha_s$. So, in the collinear parton model
we take into consideration the partonic subprocess:
\begin{equation}
g + g \to g + g.
\end{equation}
In the $k_T$-factorization approach we take into consideration the
subprocess with off-shell or reggeized initial gluons:
\begin{equation}
g^* + g^* \to g.
\end{equation}
We have described the CDF data \cite{1} and obtained a good
agreement between the parton model and the $k_T$-factorization
calculations based on the hard amlitudes for subprocesses (1) and
(2) with the equal values of the long-distance matrix elements
$<{\cal O}^{J/\psi}[^3S_1,8]>$ and $<{\cal O}^{\psi'}[^3S_1,8]>$
in the fragmentation functions $D_{g\to J/\psi,\psi'}(z,\mu^2)$
for the both models. The QCD evolution of the fragmentation
function is described by the homogeneous equation with a boundary
condition proportional to the delta function $\delta (1-z)$.

The results obtained in the paper within the $k_T$-factorization
approach based on the fragmentation function model differ from
results of Refs. \cite{7a,7b,7c}, which were obtained using the
gluon fusion model. The possible reasons of a disagreement will be
discuss below.

\section{NRQCD formalism}
 Within
the framework of the NRQCD, the cross section or the fragmentation
function for a quarkonium $H$ production can be expressed as a sum
of terms, which are factorized into a short-distance coefficient
and a long-distance matrix element \cite{3}:
\begin{eqnarray}
d\sigma (H)&=&\sum_n d\hat \sigma (Q\bar Q [n])<{\cal O}^H[n]>,\\
D(a\to H)&=&\sum_n D(a\to Q\bar Q[n])<{\cal O}^H[n]>.
\end{eqnarray}
Here the $n$ denotes the set of color and angular momentum numbers
of the $Q\bar Q$ pair, which production cross section is $ d\hat
\sigma (Q\bar Q [n])$ or which fragmentation function is $D(a\to
Q\bar Q[n])$. Last ones can be calculated perturbatively in the
strong coupling $\alpha_s$. Of course, in the case of production
in a hadron collision, the short-distance cross-section $d\hat
\sigma$ have to be convoluted with the parton distribution
function from the hadrons. The nonperturbative transition from the
$Q\bar Q$ state $n$ into the final quarkonium $H$ is described by
a long-distance matrix element $<{\cal O}^H[n]>$ which have to be
calculated using non perturbative methods or determined from
experimental data. The fit of the Tevatron data for the
$p_T$-spectra of $J/\psi$, $\psi'$ and $\chi_c$ charmonium states
have been done recently by different authors (see review in Ref.
\cite{10}). As it was shown in Refs. \cite{8,9}, at the large
charmonium transverse momentum ($p_T>7$ GeV) the gluon
fragmentation into color-octet state
\begin{equation}
g^* \to Q\bar Q[^3S_1,8]
\end{equation}
gives dominated contribution to the direct $J/\psi$ and $\psi'$
hadroproduction. The values of the color-octet matrix elements
under consideration are following $<{\cal
O}^{J/\psi}[^3S_1,8]>$=$4.4\cdot 10^{-3}$ GeV$^3$,
 $<{\cal O}^{\psi'}[^3S_1,8]>$=$4.2 \cdot10^{-3}$ GeV$^3$ \cite{9}.
Note, that the fit of CDF data for the $p_T$-spectrum of a direct
$J/\psi$ production
 within the framework of the fusion model in the
collinear parton model gives the another numerical value of the
long-distance matrix element $<{\cal
O}^{J/\psi}[^3S_1,8]>$=$1.2\cdot 10^{-2}$ GeV$^3$ \cite{10}. It
will be important to compare the results obtained in the
$k_T$-factorization approach based on the fusion and fragmentation
models.

\section{Fragmentation function}

The gluon fragmentation into $^3S_1$ charmonium state is
determined by the color-singlet (Fig.1,a) and the color-octet
(Fig.1,b) contributions. The previous analysis has shown that the
probability of a gluon fragmentation into the color-singlet state
is only a small part of the probability of a fragmentation into
the color-octet state \cite{8}. The leading order in the
$\alpha_s$ fragmentation functions for the transition (4) are
known and can be written at the scale $\mu^2=\mu^2_0=4m_c^2$ as
follows \cite{4}:
 \begin{equation}
D^T_{g\to J/\psi, \psi'}(z,\mu^2)=2d^T(z,\mu_0^2)<{\cal
O}^{J/\psi,\psi'}[^3S_1,8]>,
\end{equation}
\begin{equation}
D^L_{g\to J/\psi, \psi'}(z,\mu^2)=d^L(z,\mu_0^2)<{\cal
O}^{J/\psi,\psi'}[^3S_1,8]>,
\end{equation}
 where
 \begin{equation}
 d^T(z,\mu_0^2)=\frac{\pi \alpha_s(\mu_0^2)}{48 m_c^3}\delta (1-z)
 \end{equation}

 \begin{equation}
 d^L(z,\mu_0^2)=\frac{\alpha_s^2(\mu_0^2)}{8
 m_c^3}\frac{(1-z)}{z}.
 \end{equation}

 It is obviously that the probability of the gluon fragmentation
 into the longitudinally polarized charmonium is negligibly small
 and $J/\psi$ or $\psi'$ mesons have transverse polarization.

 The fragmentation functions (7) and (8) are evolved in $\mu^2$ using the
 standard homogeneous DGLAP evolution equation \cite{11}:
 \begin{equation}
\mu^2\frac{\partial D_g}{\partial
\mu^2}(z,\mu^2)=\frac{\alpha_s(\mu^2)}{2\pi}\int_z^1
\frac{dx}{x}P_{gg}(\frac{x}{z})D_g(x,\mu^2),
 \end{equation}
 where $P_{gg}(x)$ is the usual leading order gluon-gluon
 splitting function. To solve equation (9) we use the well known
 method based on Mellin transform. It is easy to obtain that
 the Mellin-momentum at the scale $\mu^2$ can be written as
 follows
 \begin{equation}
 D_g(n,\mu^2)=D_g(n,\mu_0^2) \exp \Biggl[ \frac{P_{gg}(n)}{2\pi}
 \int_{\mu_0^2}^{\mu^2}\frac{d\mu^2}{\mu^2}\alpha_s(\mu^2) \Biggr],
 \end{equation}
 where
 \begin{equation}
P_{gg}(n)=3\biggl[-2S_1(n)+\frac{11}{6}+\frac{2}{n(n-1)}+\frac{2}{(n+1)(n+2)}\biggr]-1,
 \end{equation}
 $$ S_1(n)=\sum_{j=1}^n\frac{1}{j}.$$
 In the one-loop  approximation for the running constant
 $\alpha_s(\mu^2)$ with the three active flavors ($b_0=9$) one has
 \begin{equation}
 \alpha_s(\mu^2)=\frac{4\pi}{b_0\log (\mu^2/\Lambda^2)},
 \end{equation}
 and equation (10) can be presented as follows
 \begin{equation}
D_g(n,\mu^2)=D_g(n,\mu_0^2) \exp \Biggl[
\frac{2}{b_0}P_{gg}(n)\log \biggl(\frac{\log
(\mu^2/\Lambda^2)}{\log(\mu_0^2/\Lambda^2)}\biggr) \Biggr].
 \end{equation}
 Taking into consideration that
 \begin{equation}
D_g^T(n,\mu_0^2)=\frac{\pi\alpha_s(\mu_0^2)}{24m_c^3}<{\cal
O}^{J/\psi,\psi'}[^3S_1,8]>
 \end{equation}
 we have performed inverse Mellin transform numerically using
 following rule
 \begin{equation}
D_g(z,\mu^2)\approx D_g^T(z,\mu^2)=\int_C dn z^{-n}D_g^T(n,\mu^2).
 \end{equation}
 The integration contour $C$ can be transformed such as
 \begin{equation}
D_g(z,\mu^2)=\frac{1}{\pi}\int_0^\infty dt \quad \mbox{Im}\left[
e^{i\phi}z^{-n}D_g^T(n,\mu^2)\right],
 \end{equation}
 where $n=c+t e^{i\phi}$ and $c\approx 2$, $\phi=\displaystyle{\frac{\pi}{2}}$.
In Fig.2 the obtained fragmentation function $D_{g\to
J/\psi}(z,\mu^2)$ multiplied by $10^4$ is shown at the different
$\mu^2=30, 100$ and $300$ GeV$^2$. We see that our result at the
$\mu^2=300$ GeV$^2$ agrees well with the same one obtained in Ref.
\cite{12}. In a stage of convoluting of the fragmentation function
$D_{g\to J/\psi}(z,\mu^2)$ with the partonic cross section for the
subprocesses (1) or (2) we will use the following definition for
the variable $z$:
\begin{equation}
z=\frac{E_{\psi}+|\vec p_{\psi}|}{2E_g}
\end{equation}
Thus we consider that the massless parton fragments into the
massive meson. As it will be seen the definition (18) is more
correct at a not so large $p_T$ than the massless one between the
parton 4-momentum and the meson 4-momentum
\begin{equation}
p_{\psi}^{\mu}=z p_g^{\mu}.
\end{equation}
We suggest also that meson has a small transverse momentum
respectively initial gluon jet and we approximately can accept
that in the laboratory frame
\begin{equation}
\eta_{\psi}\cong \eta_g,
\end{equation}
where $\eta_{\psi}, \eta_g$ are the meson and gluon
pseudorapidities.

 Within the framework of the fragmentation model, the meson production
cross section and the relevant gluon production cross section are
connected as follows
\begin{equation}
\hat \sigma(gg\to J/\psi X)=\int dz D_{g\to J/\psi}(z,\mu^2)\hat
\sigma(gg\to gg)
\end{equation}
or
\begin{equation}
\hat \sigma(g^*g^*\to J/\psi X)=\int dz D_{g\to
J/\psi}(z,\mu^2)\hat \sigma(g^*g^*\to g).
\end{equation}

\section{Leading order hard amplitudes}
The squared amplitude for the partonic process (1) is well known
and it can be presented as follows
\begin{equation}
\overline{|M(gg\to gg)|^2}= 18\pi^2 \alpha_s^2
    \frac{(\hat s^4+\hat t^4+\hat u^4)(\hat s^2+\hat t^2+\hat u^2)}{(\hat s\hat t\hat
    u)^2},
\end{equation}
where $\hat s$, $\hat t$ and $\hat u$ are usual Mendelstam
variables.

There two approaches for calculation of a partonic amplitude  for
the subprocess (2) in the $k_T$-factorization approach
\cite{5b,5d}. The effective Feynman rules for processes with
off-shell gluons were suggested in Ref. \cite{5b}. The special
trick is a choice  of the initial gluon polarization 4-vector as
follows
\begin{equation}
\varepsilon^{\mu}(k_T)=\frac{k_T^{\mu}}{|\vec k_T|}.
\end{equation}
In Ref. \cite{5d} the initial gluons are considered as reggeons
(reggeized gluons) and the effective Reggeon-Reggeon-Gluon vertex
function was obtained
\begin{eqnarray}
C^{\lambda}(k_1,k_2)=-(k_1-k_2)^{\lambda}+P_1^{\lambda}\left(
\frac{k_1^2}{(kP_1)}+2\frac{(kP_2)}{(P_1P_2)}\right)
-P_2^{\lambda}\left(
\frac{k_2^2}{(kP_2)}+2\frac{(kP_1)}{(P_1P_2)}\right),
\end{eqnarray}
where $P_1=\frac{\sqrt{s}}{2}(1,0,0,1)$ and
$P_1=\frac{\sqrt{s}}{2}(1,0,0,-1)$ are the colliding protons
4-momenta, $k_1=x_1P_1+k_{1T}$ and $k_2=x_2P_2+k_{2T}$ are the
initial gluons 4-momenta, $k_T=(0,\vec k_T,0)$, $k=k_1+k_2$ is the
final real gluon 4-momentum. It is easy to show that vertex
function $C^{\lambda}(k_1,k_2)$ satisfies the gauge invariance
condition $(k_1+k_2)_{\lambda}C^{\lambda}(k_1,k_2)=0$.

Omitting the color factor $f^{abc}$ we can write the amplitude of
the subprocess (2) accordingly Ref.\cite{5b} as follows
\begin{equation}
{\cal
M}=-g\varepsilon^{\lambda}(k)\frac{k_{1T}^{\mu}k_{2T}^{\nu}}{|\vec
k_{1T}||\vec k_{2T}|}\left[(k+k_1)_\nu
g_{\lambda\mu}+(-k_1+k_2)_\lambda g_{\mu\nu}+(-k_1-k)_\mu
g_{\nu\lambda}\right].
\end{equation}
We have obtained after simple transformations that
\begin{equation}
{\cal M}=-\frac{g\varepsilon^{\lambda}(k)}{2|\vec k_{1T}||\vec
k_{2T}|}x_1x_2s\tilde{C_{\lambda}}(k_1,k_2),
\end{equation}
where
\begin{equation}
\tilde{C^{\lambda}}(k_1,k_2)=-(k_1-k_2)^{\lambda}+\frac{2P_1^{\lambda}}{x_2s}\left(
k_1^2+x_1x_2s\right) -\frac{2P_2^{\lambda}}{x_1s}\left(
k_2^2+x_1x_2s\right)=C^{\lambda}(k_1,k_2). \end{equation}
 Such a
way, the approaches \cite{5b,5d} are equivalent and give the equal
answer for the squared vertex function and amplitude:
\begin{equation}
C^{\lambda}(k_1,k_2)C_{\lambda}(k_1,k_2)=-\frac{4k_1^2k_2^2}{x_1x_2s},
\end{equation}
and
\begin{equation}
\overline{|M(g^*g^*\to g)|^2}=\frac{3}{2}\pi\alpha_s\vec p_T^2,
\end{equation}
where $\vec p_T^2=(\vec k_{1T}+\vec k_{2T})^2=x_1x_2s$, $\vec p_T$
is the transverse momentum of the final gluon.

\section{Cross sections for the process $pp\to J/\psi (\psi')X$}
 In the conventional collinear parton
model it is suggested that hadronic cross section, in our case,
$\sigma (pp\to J/\psi X, s)$, and the relevant partonic cross
section $\hat \sigma(gg\to J/\psi X,\hat s)$ are connected as
follows
\begin{equation}
\sigma^{PM}(pp\to J/\psi,s)=\int dx_1 \int dx_2
G(x_1,\mu^2)G(x_2,\mu^2)\hat \sigma(gg\to J/\psi,\hat s),
\end{equation}
where $\hat s=x_1x_2s$, $G(x,\mu^2)$ is the collinear gluon
distribution function in a proton, $x_{1,2}$ are the fraction of a
proton momentum, $\mu^2$ is the typical scale of a hard process.
The $\mu^2$ evolution of the gluon distribution $G(x,\mu^2)$ is
described by DGLAP evolution equation \cite{11}.

In the $k_T$-factorization approach hadronic and partonic cross
sections are related by the following condition \cite{5a,5b,5c}:
\begin{eqnarray}
&&\sigma^{KT}(pp\to J/\psi X)=\int \frac{dx_1}{x_1}\int d{\vec
k_{1T}^2}\int \frac{d\phi_1}{2\pi} \Phi(x_1,\vec k_{1T}^2,\mu^2)\\
\nonumber &&\int \frac{dx_2}{x_2}\int d{\vec k_{2T}^2}\int
\frac{d\phi_2}{2\pi} \Phi(x_2,\vec k_{2T}^2,\mu^2)\hat \sigma(g^*
g^*\to J/\psi X, \hat s),
\end{eqnarray}
where $\hat \sigma(g^* g^*\to J/\psi X, \hat s)$ is the $J/\psi$
production cross section on off-shell gluons,
$k_1^2=k_{1T}^2=-\vec k_{1T}^2$, $k_2^2=k_{2T}^2=-\vec k_{2T}^2$,
$\hat s=x_1x_2s-(\vec k_{1T}+\vec k_{2T})^2$, $\phi_{1,2}$ are the
azimuthal angles in the transverse $XOY$ plane between vectors
$\vec k_{1T}(\vec k_{2T})$ and the fixed $OX$ axis ($\vec p_{\psi}
\in XOZ$). The unintegrated gluon distribution function
$\Phi(x_1,\vec k_{1T}^2,\mu^2)$ satisfies the BFKL evolution
equation \cite{13}.

Our calculation in the parton model is down using the GRV LO
\cite{14} and CTEQ5L \cite{15} parameterizations for a collinear
gluon distribution function $G(x,\mu^2)$. In case of the
$k_T$-factorization approach we use the following
parameterizations for an unintegrated gluon distribution function
$\Phi(x_1,\vec k_{1T}^2,\mu^2)$: JB by Bluemlein \cite{16}, JS by
Jung and Salam \cite{17}, KMR by Kimber, Martin and Ryskin
\cite{18}. The direct comparison between different
parameterizations as functions of $x$, $\vec k_T^2$ and $\mu^2$
was presented in paper \cite{7}.

The doubly differential cross sections for the process $pp\to
J/\psi (\psi')X$ can written as follows
\begin{equation}
\frac{d\sigma^{PM}}{d\eta_{\psi}dp_{\psi T}}=\int dx_1\int dz
G(x_1,\mu^2)G(x_2,\mu^2)D_{g\to
\psi}(z,\mu^2)\frac{p_{gT}E_g}{E_{\psi}}\frac{\overline{|M(gg\to
gg)|^2}}{8\pi x_1x_2s(u+x_1s)},
\end{equation}
where
$$u=-\sqrt{s}(E_g-p_{gz}), \quad t=-\sqrt{s}(E_g+p_{gz}), \quad
x_2=-\frac{x_1 t}{u+x_1s}, \quad x_{1,min}=-\frac{u}{s+t},$$
$$ \hat t=x_1t, \quad \hat u=x_2u, \quad \hat s=x_1x_2s;$$
\begin{equation}
\frac{d\sigma^{KT}}{d\eta_{\psi}dp_{\psi T}}=\int dz\int
d\phi_1\int d\vec k_{1T}^2 \Phi(x_1,\vec
k_{1T}^2,\mu^2)\Phi(x_2,\vec k_{2T}^2,\mu^2)D_{g\to
\psi}(z,\mu^2)\frac{E_g}{p_{gT}E_{\psi}}\frac{\overline{|M(g^*g^*\to
g)|^2}}{x_1 x_2 s},
\end{equation}
where
$$\vec k_{2T}=\vec p_{gT}-\vec k_{1T}, \quad
x_1=\frac{E_g+p_{gz}}{\sqrt{s}}, \quad
x_2=\frac{E_g-p_{gz}}{\sqrt{s}}.$$ The energy $(E_g)$ of a
fragmenting gluon and the energy of a final meson $(E_{\psi})$ are
related as follows
$$ E_g=\frac{E_{\psi}+|\vec p_{\psi}|}{2z}, \quad |\vec p_{\psi}|=\frac{p_{\psi T}}{\sin
(\theta_{\psi})},
\quad p_{gz}=E_g\cos (\theta_{\psi}), \quad \theta_{\psi}=2~
\mbox{arctg}(\exp(-\eta_{\psi})).$$

\section{The results}
We compare our predictions with the CDF data \cite{1} for
unpolarized direct $J/\psi$ and $\psi'$ production. The direct
$J/\psi$ cross section does not include contributions from $B$
meson decays into $J/\psi$ as well as from $\chi_c$ radiative
decays. For direct $\psi'$ production, indirect contribution from
$B$ meson decays are removed. Our results obtained in the
collinear parton model do not depend on a choice of the
parameterization for the gluon distribution function and coincide
approximately  ($\pm$10-20\%) to the results obtained in Ref.
\cite{9} using a similar approch. We can see in Figs. 3 and 4 that
the curves denoted as "GRV", which were  obtained in the collinear
parton model, are below the data especially at the small $p_T$,
were the gluon fusion into $^1S_0$ and $^3P_J$ color octet states
is dominant \cite{9,10}.

The curves which were obtained in the $k_T$-factorization approach
strongly depend on a choice of the unintegrated gluon distribution
function. In the region of a large $p_T$, where the fragmentation
approach is more adequate, the results obtained with JB \cite{16}
and KMR \cite{18} parameterizations coincide well. However, JS
parameterization \cite{17} predicts the values smaller by factor
2, which are near the values obtained in the collinear parton
model with the GRV gluon distribution function. Such a way, the
$p_T$ spectra of direct $J/\psi$ and $\psi'$ mesons obtained in
the collinear parton model and in the $k_T$-factorization approach
are approximately coincide. Note, that our conclusions disagree
with the previous results obtained in Refs. \cite{7a,7b,7c} using
the fusion model.  Opposite our result, the fit of the CDF data
accordingly \cite{7a,7b,7c} needs strong suppression (in 10-30
times) for the long-distance matrix elements $<{\cal
O}^{J/\psi,\psi'}[^3S_1, \underline{8}]>$ to compare the values
obtained in the collinear parton model.

There are several reasons of a such disagreement. At first, we
used the fragmentation functions, which take into account
effectively high order corrections via the DGLAP evolution
equation. At second, in Refs. \cite{7a,7b,7c} the argument $\mu^2$
of the strong coupling constant $\alpha_s(\mu^2)$ is equal to
$\vec k_{1T}^2$ or $\vec k_{2T}^2$, our choice is $\mu^2=\vec
p_T^2+4m_c^2$. This fact gives the additional  factor 3 in a cross
section \cite{7c}. At third, in Refs.\cite{7a,7b} the KMS\cite{22}
parameterization for an unintegrated gluon distribution function
was used. We have shown (Figs. 2 and 3) that the difference
between predictions based on the different parameterizations may
be about factor 2-3.

The obtained results for the direct unpolarized $J/\psi$ and
$\psi'$ hadroproduction as well as our previous results for the
$J/\psi$ photoproduction at HERA energies \cite{7} show that the
predictions for spectra on $p_T$ with the LO in $\alpha_s$ hard
amplitudes in the $k_T$-factorization approach coincide well to
the predictions with the NLO in $\alpha_s$ hard amplitudes in the
collinear parton model. The NLO in $\alpha_s$ calculation for the
$J/\psi$ photoproduction cross section was performed in Ref.
\cite{23}.

It is obviously that the NLO hard subprocess for a gluon
production in the $k_T-$factorization approach with off-shell
initial gluons is the LO subprocess used in the collinear parton
model with on-shell initial gluons:
\begin{equation}
g^*+g^*\to g+g.
\end{equation}
An amplitude of the subprocess (34) has infrared singularities
even at the large value of $p_T$ for the final gluon which
fragments into a meson. Opposite, in the collinear parton model
the both gluons to be hard in similar case. The procedure of a
calculation of an amplitude for the subprocess (34) in the
$k_T$-factorization approach has been suggested in Ref.\cite{5d},
were initial gluons are  considered as reggeons and the infrared
divergencies are removed. It should be interesting to calculate
$J/\psi (\psi')$ production cross section using the results of
Ref. \cite{5d}. This study is in progress.

\subsection*{Acknowledgements}
 We thank H.~Jung for valuable information on
parameterizations for an unintegrated gluon structure function and
B.~Kniehl, O.~Teryaev and L.~Szymanowski for useful discussion of
the obtained results. One of us (D.V.) thanks IHEP Directorate and
A.M.~Zaitsev for a kind hospitality during his visit in Protvino
where the part of this work was done. The work was supported by
the Russian Foundation for Basic Research under Grant 02-02-16253.

\newpage

\begin{figure}[ht]
\begin{center}
\includegraphics[width=.6\textwidth, clip=]{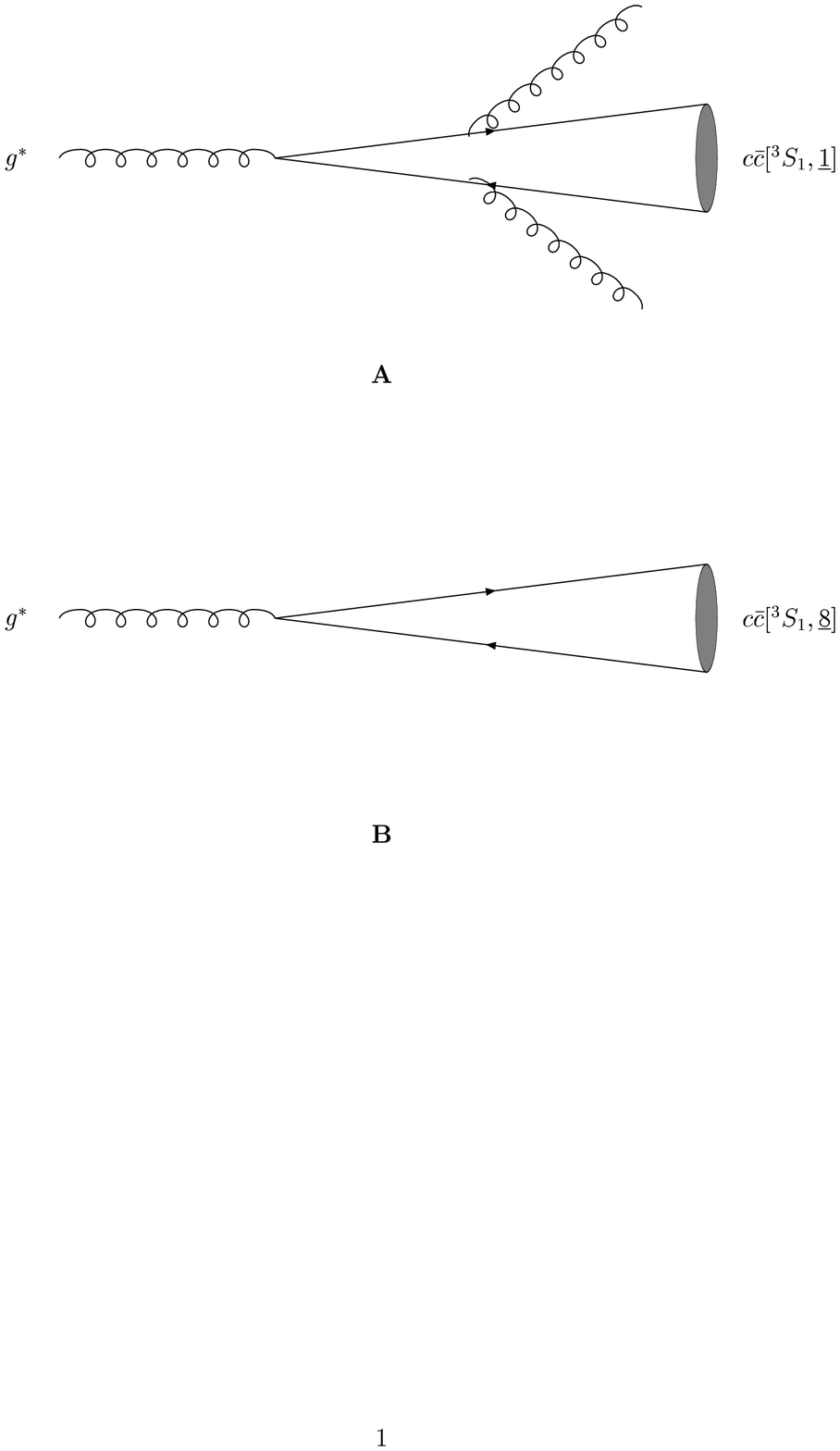}
\end{center}
\caption[]{ Diagrams used for description LO in the $\alpha_s$
fragmentation $g^*\to J/\psi gg$ (color-singlet state, A) and
$g^*\to J/\psi$ (color-octet state, B).}
\end{figure}

\begin{figure}[ht]
\begin{center}
\includegraphics[width=.6\textwidth, clip=]{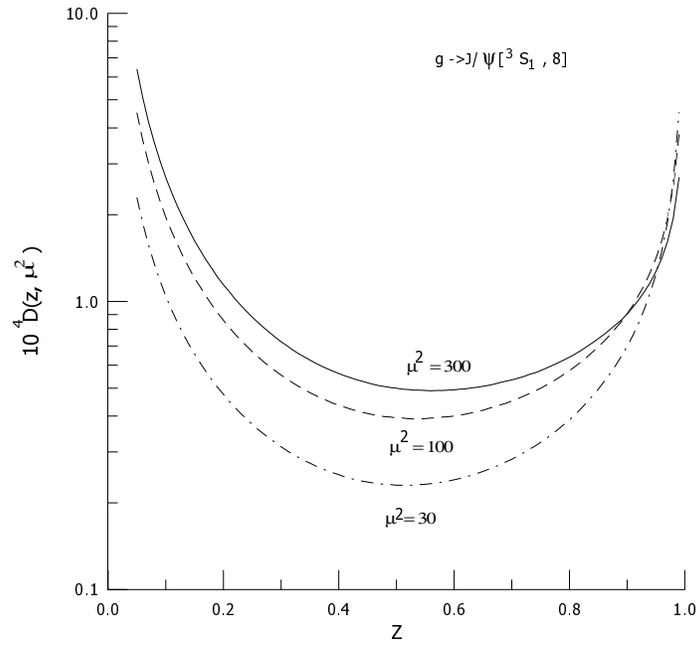}
\end{center}
 \caption[]{ $D_{g\to J/\psi}(z,\mu^2)$ at the
$\mu^2$=30, 100
 and 300 GeV$^2$ as a function of z.}
\end{figure}

\begin{figure}[ht]
\begin{center}
\includegraphics[width=.6\textwidth, clip=]{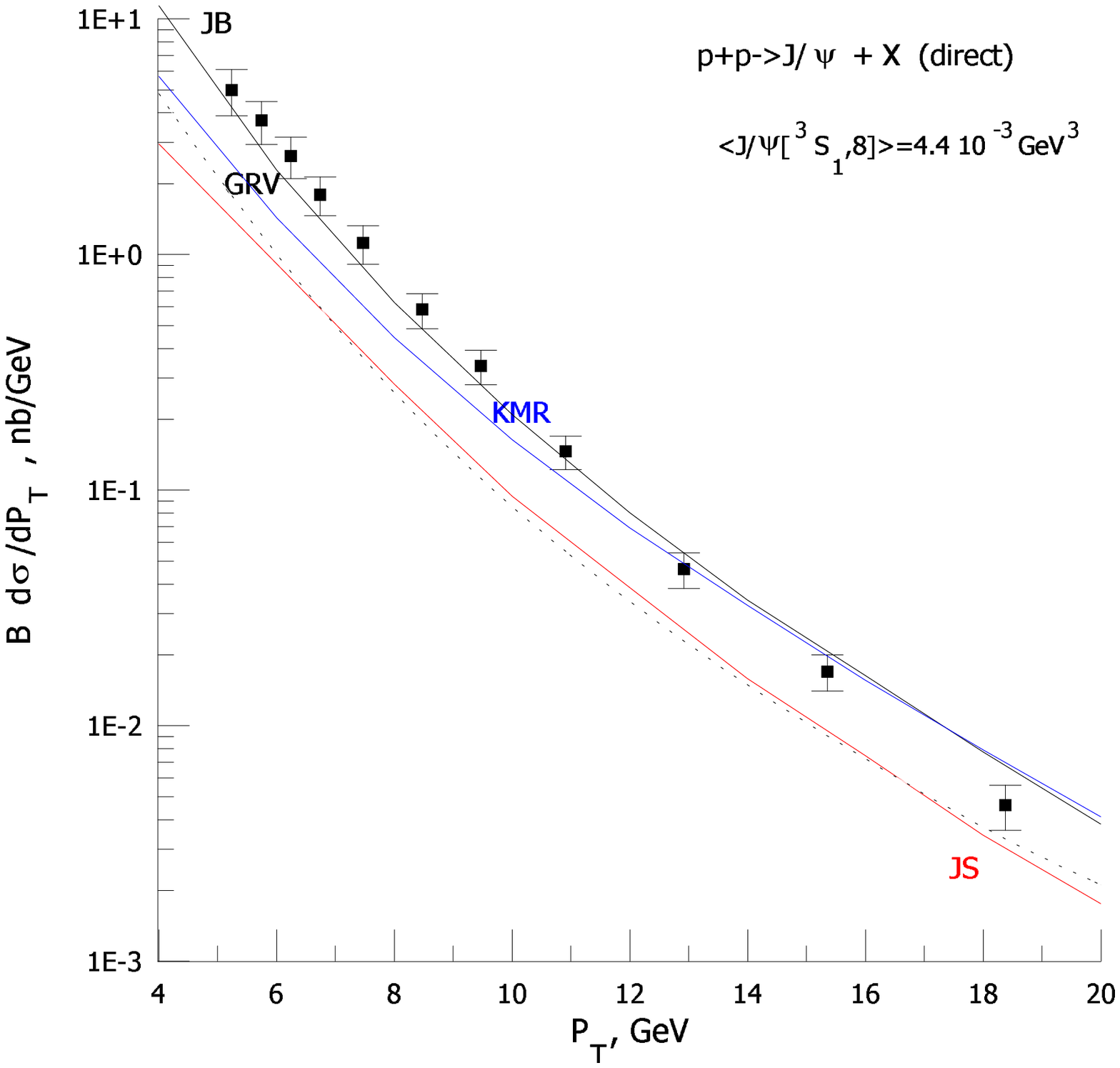}
\end{center}
\caption[]{ The spectrum of direct $J/\psi$ on $p_T$ at the
$\sqrt{S}$=1800 GeV and  $|\eta|<0.6$. The data points are from
[1]. The $B$ is the $J/\psi$ lepton branching.}
\end{figure}

\begin{figure}[ht]
\begin{center}
\includegraphics[width=.6\textwidth, clip=]{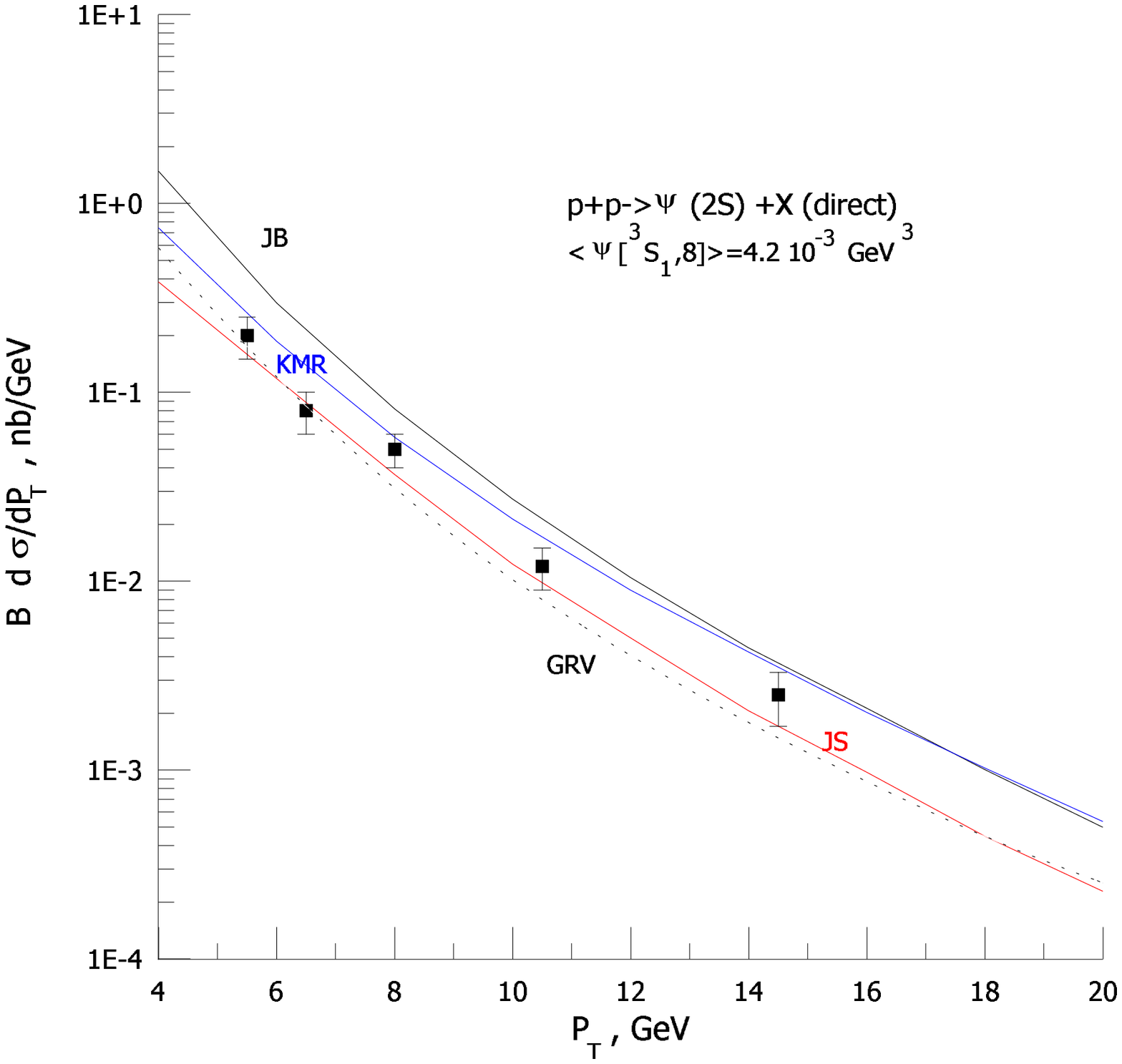}
\end{center}
\caption[]{ The spectrum of direct $\psi'$ on $p_T$ at the
$\sqrt{S}$=1800 GeV and  $|\eta|<0.6$. The data points are from
[1]. The $B$ is the $\psi'$ lepton branching.}

\end{figure}
\end{document}